\documentclass[aps,prd,twocolumn,groupedaffiliation,nofootinbib]{revtex4-1}
\pdfoutput=1
\usepackage{graphicx}
\usepackage{tikz}
\usepackage{graphicx, epsfig}
\usepackage{color,soul}
\usepackage{mathrsfs}
\usepackage{amsmath}
\usepackage{amssymb}
\usepackage{hyperref}
\usepackage{tabularx}
\newcommand{\ben}{\begin{enumerate}}
\newcommand{\een}{\end{enumerate}}
\newcommand{\spa}{\phantom{s}}
\newcommand{\bea}{\begin{eqnarray}}
\newcommand{\eea}{\end{eqnarray}}
\newcommand{\be}{\begin{equation}}
\def\bel#1{\begin{equation} \label{#1}}
\newcommand{\ee}{\end{equation}}
\newcommand{\bi}{\begin{itemize}}
\newcommand{\ei}{\end{itemize}}
\newcommand{\ba}{\begin{align}}
\newcommand{\ea}{\end{align}}
\def\prd #1 #2 #3 {#1, Phys.~Rev.~D., {\bf #2}, #3}
\def\bel#1{\begin{equation} \label{#1}}

\def\b{\beta}

\def\mc{\mathcal}
\def\be{\begin{equation}}
\def\ee{\end{equation}}
\def\bea{\begin{eqnarray}}
\def\eea{\end{eqnarray}}

\def\ltap{\ \raise.3ex\hbox{$<$\kern-.75em\lower1ex\hbox{$\sim$}}\ }
\def\gtap{\ \raise.3ex\hbox{$>$\kern-.75em\lower1ex\hbox{$\sim$}}\ }
\def\gl{\ \raise.5ex\hbox{$>$}\kern-.8em\lower.5ex\hbox{$<$}\ }
\def\roughly#1{\raise.3ex\hbox{$#1$\kern-.75em\lower1ex\hbox{$\sim$}}}

\def\spa{\phantom{a}}
\def\pref#1{(\ref{#1})}

\def\cv{{{\cal{V}}}}

\newcommand{\comments}[1]{}
\usepackage{color}
\definecolor{cblue}{RGB}{100,5,255}
\definecolor{cred}{RGB}{255,50,40} 
\definecolor{cgreen}{RGB}{40,255,40} 
\definecolor{cmagenta}{RGB}{139,0,139}
     
\begin{document}

\bigskip
%
\title{Confronting K\"ahler moduli inflation with CMB data }
%
%

\author{Sukannya Bhattacharya}
\email[]{sukannya.bhattacharya@saha.ac.in}
\affiliation{Theory Divison, Saha Institute of Nuclear Physics, HBNI,1/AF Bidhannagar, Kolkata- 700064, India }
\author{Koushik Dutta}
\email[]{koushik.dutta@saha.ac.in}
\affiliation{Theory Divison, Saha Institute of Nuclear Physics, HBNI,1/AF Bidhannagar, Kolkata- 700064, India }
\author{Mayukh Raj Gangopadhyay}
\email[]{mayukh.raj@saha.ac.in}
\affiliation{Theory Divison, Saha Institute of Nuclear Physics, HBNI,1/AF Bidhannagar, Kolkata- 700064, India }
\author{Anshuman Maharana}
\email[]{anshumanmaharana@hri.res.in}
\affiliation{Harish Chandra Research Institute, HBNI, Chattnag Road, Jhunsi, Allahabad -  211019, India.}


\begin{abstract}

\noindent  
In models of inflation obtained from string compactification, moduli vacuum misalignment leads to an epoch in the post-inflationary history of the universe when the energy density is dominated by cold moduli particles. This effect leads to a modification in the number of $e$-foldings ($N_{\rm pivot}$) between horizon exit of the CMB modes and the end of inflation. Taking K\"ahler moduli inflation as a prototype, the shift in $e$-foldings turns out to be a function of the model parameters which also determines the inflationary observables. We analyse this scenario numerically using publicly available {\sc{ModeChord}} and {\sc{CosmoMC}} with the latest \emph{Planck+BICEP2/Keck array} data to constrain the model parameters and $N_{\rm pivot}$. {In light of the present and future precision data, the results show the importance of careful consideration of any post-inflationary non-standard epoch, as well as of  the effects of reheating.}

\end{abstract}


\maketitle
%
%
%




\section{Introduction}
\label{Seckmi}

The inflationary paradigm has provided an explanation for the observed  spectrum and inhomogeneities in the Cosmic Microwave Background (CMB). Intense efforts are on to probe the CMB for tensor modes and non-Gaussianity. Observations of these will provide further credence to the theory. At the same time, programs to measure the spectral tilt of scalar modes $(n_s)$  with greater accuracy are being planned. The  ground based CMB-S4 experiment  \cite{Aba}, the LiteBIRD satellite \cite{Matsumura:2013aja}, and the CORE satellite \cite{core}, if approved, can significantly reduce the uncertainty in the measurement of the spectral tilt (projected uncertainty $\sigma(n_s) \sim 0.002$ at the $1$-$\sigma$ level). In this light, it is important to develop  tools to extract accurate predictions for inflationary models.

The most commonly  adopted method to constrain models of inflation is to express the primordial perturbations in terms of empirical parameters such as $A_s$ (the strength of the power spectrum), $n_s$ (the scalar tilt), $r$ (the tensor to scalar ratio), $f_{nl}$ (parametrising the non-Gaussianity) etc. The most likely values of the empirical  parameters are determined by evolving the initial fluctuations (as expressed in terms of  these parameters) and then comparing with the observed CMB fluctuations. Given a model of inflation, its theoretical prediction for the empirical parameters are computed as a function of model parameters, and a model is considered successful if the predicted values match the constraints on the empirical parameters from the observations. This indeed is the general procedure followed  by the Planck collaboration to obtain constraints on several inflation models\cite{planck2015}. {In this methodology, it is important to remember that the observables need to be calculated at certain e-folds back ($N_{\rm pivot}$) from the end of inflation when CMB modes go outside the horizon, and $N_{\rm pivot}$ crucially depends on the post-inflationary cosmic history including the epoch of reheating, see for e.g \cite{Liddle:2003as}.}
   
If one is interested in confronting a particular model of inflation with data, then a robust  approach can be taken, as developed in \cite{Jmartin3, Martin:2010kz, modecode}. One takes the coefficients of the inflaton  potential and the parametrisation of the reheating epoch as the `model inputs'. Observational predictions are examined directly in terms of the coefficients of the potential; estimates and errors for the coefficients of the potential are directly obtained. One of the ways, this can be achieved is by making use of {\sc{ModeChord}}\footnote{{\sc{ModeChord}} is publicly available at \href{http://www.modecode.org}{www.modecode.org}}\cite{modecode} which  provides a numerical evaluation of the inflationary perturbation spectrum  (even without relying on the slow-roll approximation) taking the potential coefficients as input; which is then used as a plug-in for {\sc{CAMB}}\cite{cambb} and {\sc{CosmoMC}}\cite{cosmomc_main}. The parameters are then estimated using a nested sampling method\cite{multinest}.
{The importance of reheating effects in constraining inflation models  using current cosmological data was first discussed in \cite{Martin:2010kz}, and was subsequently applied to the WMAP data in \cite{Martin:2010hh} and to the Planck data in \cite{Martin:2013nzq}.}

The  slow-roll conditions require the inflationary potential to be flat in Planck units. Any inflation model is sensitive to ultraviolet degrees of freedom, and therefore, inflation models should be embedded in ultraviolet complete theories. String theory being our best hope for an ultraviolet complete theory, inflationary models obtained from string theory deserve to be analysed in detail. This is best achieved by following the approach of \cite{modecode}.  Inflationary model building  has been extensively studied in string theory (see e.g \cite{macb}).  

There is also a more pragmatic reason to analyse models obtained from string theory using this approach. String compactifications typically have moduli fields, which are massless scalars with interactions of gravitational strength.  A generic feature of models of inflation constructed in string theory (and supergravity) is vacuum misalignment of moduli. This leads to an epoch in the post-inflationary history in which the energy density of the universe is dominated by cold moduli particles  (see e.g. \cite{Kane} for a recent review). The presence of such an epoch changes the number of e-foldings of the universe between horizon exit of the pivot mode and the end of inflation ($N_{\rm pivot}$). Since the primordial perturbations are determined by  local properties of the inflaton potential at the time of horizon exit, $N_{\rm pivot}$ has an effect on inflationary predictions. Recent work has shown that it is important to incorporate this effect while confronting string models with precision cosmological data \cite{BLH,D}; an accurate determination of $N_{\rm pivot}$ is essential. Furthermore,  $N_{\rm pivot}$ has dependence on the model parameters that also determine inflationary observables. Given this, a complete numerical methodology using {\sc{ModeChord+CosmoMC}} \cite{modecode, cosmomc_main} is carried out in this work. In this paper, we will take  K\"ahler moduli inflation \cite{Kahler}, \cite{Lee:2010tk} as a prototype and analyse it using {\sc{ModeChord}}.  Vacuum misalignment  in the model was studied in \cite{BLH}, where the effect of this epoch on $N_{\rm pivot}$ was determined. {We note that K\"ahler moduli inflation in light of Bayesian model selection was discussed in  \cite{Martin:2014nya, Martin:2013tda, m3}, they key difference between the present analysis and that of \cite{Martin:2014nya, Martin:2013tda, m3} is incorporation the effects of  the fact that the  duration of the epoch of modulus domination  in the post-inflationary history depends on the model parameters. }
     
This paper is organised as follows. In the next section, we first  review some basic aspects of K\"ahler moduli inflation relevant for us. We then mention the duration of modulus dominated epoch at the end of inflation $N_{\rm mod}$ for the model. $N_{\rm mod}$ shifts the $N_{\rm pivot}$ from its usual number. In Sec.~\ref{results} we discuss the methodology for analysing the model parameters and how the required modification in $N_{\rm pivot}$ can be implemented in {\sc{ModeChord}}. {In Sec.~ \ref{section_GRH}, we analyse and discuss the results for Generalised Reheating (GRH) scenario where $N_{\rm pivot}$ is varied between $20$ and the number corresponding to the instantaneous reheating case. This analysis is independent of average equation of state parameter $w_{\rm re}$ during reheating. The case for specific values of $w_{\rm re}$ with the requirement of $T_{\rm re} > T_{BBN}$ is analysed in Sec.~\ref{general_w}}. We conclude in Sec.~\ref{conclusions}. 

  
\section{A brief review of K\"ahler moduli inflation}
\label{Seclvs}


 We begin by briefly reviewing K\"ahler moduli inflation, the reader  should consult \cite{Kahler} for further details. K\"ahler moduli inflation is set in the  Large Volume Scenario (LVS) for moduli stabilisation \cite{LVS,LVS2} of IIB flux compatifications \cite{GKP}. The complex structure moduli of the underlying Calabi-Yau are stabilised by fluxes.
The simplest models of LVS  are the ones in which the volume of the Calabi-Yau takes  the Swiss-cheese form: $\cv = \alpha \bigg( \tau_1^{3/2} - \sum_{i=2}^{n} \lambda_i \tau_i^{3/2} \bigg)$  \cite{LVS,LVS2}. Note that  the overall volume is set  by $\tau_1$; the  moduli $\tau_2 ,..., \tau_n$ are blow-up modes and correspond to the size of the holes in the compactification. Incorporating the non-perturbative effects in the superpotential, the leading $\alpha'$ correction to the K\"ahler potential and an uplift term (so that a nearly Minkowski vacuum can be obtained), the potential for the scalars in the regime $\cv \gg 1$ and  $\tau_1 \gg \tau_i \spa ( {\rm{for}} \spa  i >1)$ is 
%

\begin{align}
V_{\rm LVS} &= \sum_{i=2}^n { 8 (a_i A_i)^2 \sqrt{\tau_i} \over 3 \cv \lambda_i } e^{-2 a_i \tau_i}-\sum_{i=2}^{n} { 4 a_i A_i W_0 \over \cv^2} \tau_i e^{-a_i \tau_i}\nonumber \\ 
&+ {3 \hat\xi  W_0^2 \over 4 \cv^3 } + { D  \over \cv^{\gamma} }.
\label{total}
\end{align}
%
Here $A_i, a_i$ are the pre-factors and coefficients in the exponents of the non-perturbative terms in the superpotential and
$W_{0}$ is the  vacuum expectation value of flux superpotential. The uplift term is  $V_{\rm up} = { D  \over \cv^{\gamma} }$ with $D > 0$, $1\leq\gamma\leq 3$  (see \cite{kklt, dil, mag1, mag2, rum} for mechanisms that can lead to such as term).

In K\"ahler moduli inflation, one of the blow-up moduli ($\tau_n$) acts as the inflaton. For simplicity, we will focus on Calabi-Yau's of the Swiss Cheese form (although the analysis can be carried out in more general settings,  \cite{CD}).
Inflation takes place in region $e^{a_n \tau_n} \gg \cv^2$, here the potential is well approximated by:
\begin{align}
V_{\rm inf} &= \sum_{i=2}^{n-1} { 8 (a_i A_i)^2 \sqrt{\tau_i} \over 3 \cv \lambda_i } e^{-2 a_i \tau_i}
-  \sum_{i=2}^{n-1} { 4 a_i A_i W_0 \over \cv^2} \tau_i e^{-a_i \tau_i}\nonumber \\ 
&+{3 \hat\xi  W_0^2 \over 4 \cv^3 } + { D  \over \cv^{\gamma} } - { 4 a_n A_n W_0 \over \cv^2} \tau_n e^{-a_n \tau_n}\,.
\label{vap}
\end{align}

It is exponentially flat in the inflaton direction $(\tau_n)$. The  other directions ($\cv, \tau_i$ with $i = 2, .. ,n-1$) in field space are heavy during inflation. Integrating out the heavy  directions and canonically normalising the inflaton (we denote the canonically normalised field by $\sigma$), one finds its potential (in Planck units) to be 
\begin{align}
V &=  { g_{s} \over 8 \pi} \bigg{(}V_0 - \frac{4 W_0 a_n A_n}{\mc{V}^2_{\rm in}} \left(\frac{3 \mc{V}_{\rm in} }{4 \lambda_n} \right)^{2/3}  \sigma^{4/3}\nonumber \\ 
&\times \exp \left[-a_n \left(\frac{3 \mc{V}_{\rm in}}{4 \lambda_n}\right)^{2/3} \sigma^{4/3}\right] \bigg{)},
\label{canpot}
\end{align}
where
 \bel{can}
\frac{\sigma}{M_{\rm pl}} = \sqrt{\frac{4 \lambda_n}{3 \mc{V}_{\rm in}}} \,\tau_n^{\frac{3}{4}} ~~~~{\rm with} \spa \spa V_{0} = \frac{\b W_0^2}{\mc{V}_{\rm in}^3}~.
\ee
 $\cv_{\rm in}$ is the value of the volume during inflation  and $\beta = {3 \over 2 }  \lambda_{n} a_n^{-3/2}   (\ln \cv)^{3/2}$. Phenomenological considerations put the volume at $\cv_{\rm in} \approx 10^{5} - 10^{7}$, and we will discuss cosmological constraints on it in the next section. We note that `gobal embedding' of the model (realisation in a compact Calabi-Yau with a semi-realistic Standard Model sector) was carried out in \cite{micu}.

  Vacuum misalignment and the resulting post-inflationary moduli dynamics in this model was studied in detail in \cite{BLH}, here we summarise its conclusions.
During inflation, the volume modulus gets displaced from its global minimum. The displacement of the canonically normalised field in Planck units is:
$$
  Y = 2 R \bigg( {\hat{\xi} \over 2P} \bigg)^{2/3} 
$$
with
\begin{align}
 &R = { \lambda_{n} a_n^{-3/2} \over \bigg( \sum_{i}^{n}  \lambda_{i} a_i^{-3/2}  \bigg)},\spa  P = \sum_{i}^{n}  \lambda_{i} a_i^{-3/2},\nonumber \\
 & \textrm{and} \spa  \hat{\xi} = {\chi \over 2 (2 \pi)^3 g_s^{3/2}}~\nonumber,
\end{align}
%
where $\chi$ is the Euler number of the Calabi-Yau and $g_{s}$ is the vacuum expectation value of the dilaton. For typical values of the microscopic parameters, $Y \approx 0.1 $ which is consistent with effective field theory expectations. This leads to an epoch in the post-inflationary history in which the energy density is dominated by cold particles of the volume modulus. The number of e-foldings that the universe undergoes in this epoch is \cite{BLH}
\begin{equation}
  N_{\rm mod} = {2 \over 3}  \ln \bigg( { 16 \pi a_n^{2/3}  \cv^{5/2} Y^{4} \over 10 \lambda_n (\ln \cv)^{1/2} } \bigg)~.
  \label{nmod1}
\end{equation}
The presence of this epoch reduces the number of e-foldings between horizon exit of the pivot mode and the end of inflation by an amount $\frac{1}{4}N_{\rm mod}$. The spectral index can be calculated by using the usual slow-roll formula $n_s \approx 1- 2/N_{\rm pivot}$, but at the reduced value of $N_{\rm pivot}$. For typical values of the volume $\cv \sim 10^5-10^6$ and other parameters in this model, $N_{\rm mod}$ can be calculated. 
In K\"ahler moduli inflation, the inflaton decays to relativistic d.o.f at the end of inflation, the  energy density of these relativistic degrees of freedom  becomes subdominant quickly in comparison   with the energy density of the oscillating volume modulus (which arises as a result of vacuum misalignment). For typical values of the model parameters, this epoch of between inflation and modulus domination has a small duration and hence a negligible effect on $N_{\rm pivot}$ \cite{BLH}. We  therefore neglect this epoch in our analysis. After the decay of the volume modulus, the universe has  standard thermal history.

\section{Methodology of analysis}
\label{results}
The analysis is dependent on the exact definition of the number of $e$-foldings  $N_{\rm pivot}$. The correctly measured number of $e$-foldings at the horizon exit is crucial to constrain models of inflation with an additional epoch of post-inflationary moduli domination. This analysis is devised to see how the change in predicted number of $e$-foldings during inflation due to the secondary moduli dominated epoch effects the observables and therefore constrains the model parameters. But, even though the modulus dominated epoch is considered carefully, there is inherent uncertainties with the exact value of  $N_{\rm pivot}$ due to our poor knowledge about the details of reheating/preheating at the end of modulus domination. With better understanding of several couplings between the inflation/modulus with Standard Model d.o.f, in future we will possibly find the total duration of the thermalization process  $N_{\rm re}$ with its average equation of state parameter $w_{\rm re}$. Until this is available, it is practical to consider the reheating parameters $w_{\rm re}$ and $N_{\rm re}$ as  variables when analysing the model in light of the recent CMB data.  

The analysis is carried out using the publicly available CosmoMC \cite{cosmomc_main} and {\sc{ModeChord}} \cite{modecode} plugged together through Multinest \cite{multinest}. Given a typical model of inflation, {\sc{ModeChord}} numerically computes the primordial scalar and tensor power spectra. These primordial spectra are fed to the CAMB in the CosmoMC package with the help of the plug-in software Multinest to evolve through transfer functions. The theoretically calculated perturbations at the CMB redshift is then compared to the observed fluctuations using CosmoMC. CosmoMC is a multi-dimensional Markov Chain Monte Carlo simulator which in this case compares the $C_l$ values computed numerically for the given inflationary model with the observed $C_l$ values by Planck and BICEP-Keck array \cite{keck,commander,bkplanck}. In general, all the model parameters of inflation and late time cosmological parameters (e.g. $\Omega_b$, $\Omega_c$, $\theta$ and $\tau$) are variables in this ModeChord+CosmoMC set up. In addition, number of $e$-folds of inflation can also be set as a variable due to our lack of knowledge of the (p)reheating epoch. In this work, we have varied all the late time cosmological parameters in the six-parameter $\Lambda CDM$ model as well as the number of $e$-folds  during inflation. The ranges of the inflationary model parameters which are varied in the simulation are chosen carefully and and particularly the ranges for the reheating parameters are explained in the following paragraphs.

For the generic inflation scenario with instantaneous reheating (IRH) where the universe thermalizes instantly after inflation and makes quick transition to the radiation dominated epoch, the number of $e$-foldings at the pivot scale is given by \cite{modecode}:
\begin{equation}
 N^{\rm IRH}_{\rm pivot} = 55.75 - \log \bigg[ {10^{16} \rm Gev \over V^{1/4}_{\rm pivot}} \bigg] + \log \bigg[ { V^{1/4}_{\rm pivot} \over V^{1/4}_{\rm end}} \bigg].
 \label{standard_irh}
\end{equation}
Here, $V_{\rm pivot}$ is the value of the inflation potential at which the pivot scale leaves the horizon and $V_{\rm end}$ is the potential at the end of inflation. From the observational upper limit of the strength of the gravitational wave ($r < 0.11$ \cite{planck2015}), the second term in the above equation is negative, whereas the third term is positive definite, but it can be very small for observationally favoured flat inflaton potential. In the usual implementation of {\sc{ModeChord}}, the cosmological perturbations are evaluated without assuming slow-roll conditions, and the best-fit potential parameters can be estimated using CosmoMC. But this also requires that  the uncertainties associated with reheating are accounted for, and this can be done by varying $N_{\rm pivot}$  between $ 20 < N_{\rm pivot} <  N^{\rm IRH}_{\rm pivot}$. This is termed as the general reheating (GRH) scenario  \cite{modecode}. The upper limit is motivated from the assumption that the average dilution of energy density during the reheating epoch is not faster than radiation, i.e $w_{\rm re} \leq 1/3$. The lower limit comes from the requirement that at the end of inflation, all the cosmologically relevant scales are well outside of the horizon.  The shortcoming of this approach is that the reheating scenarios with $w_{\rm re} > 1/3$ are not considered; the possibility that $N_{\rm pivot}$ can be above $N_{\rm IRH}$ is excluded in this analysis.

If there is an epoch of moduli domination in the post-inflationary history, then Eq.~\ref{standard_irh} gets modified. For  K\"ahler moduli inflation $N_{\rm mod}$ is given by Eq.~\ref{nmod1}, and in this case $N^{\rm IRH}_{\rm pivot}$ is:
\begin{align}
N^{\rm IRH}_{\rm pivot} &= 55.75 - \log \bigg[ {10^{16} \rm Gev \over V^{1/4}_{\rm pivot}} \bigg] + \log \bigg[ { V^{1/4}_{\rm pivot} \over V^{1/4}_{\rm end}} \bigg] \nonumber \\ 
&- {1 \over 6}  \ln \bigg( { 16 \pi a_n^{2/3}  \cv^{5/2} Y^{4} \over 10 \lambda_n (\ln \cv)^{1/2} } \bigg)~.
\label{kahler_irh}
\end{align}
Note the additional dependence on the model parameters that arises from the last term in Eq.~\eqref{kahler_irh}. In our analysis for $ -1/3 <w_{\rm re} \leq 1/3$, we will vary $N_{\rm pivot}$ between 20 and $N^{\rm IRH}_{\rm pivot}$ given by Eq.~\eqref{kahler_irh}.

In general, $N_{\rm pivot}$ is determined by  $N^{\rm IRH}_{\rm pivot}$  (as determined by equation \pref{kahler_irh}), $w_{\rm re}$ and  $N_{\rm re}$:
\bel{nppp}
   N_{\rm pivot} =  N^{\rm IRH}_{\rm pivot}  - {1 \over 4} ( 1 - 3 w_{\rm re} ) N_{\rm re}. 
\ee
%
%
%
The most general reheating case for the modulus can be treated with considering $-1/3 < w_{\rm re} < 1$, where the upper bound comes from the positivity conditions in general relativity. 
The GRH analysis as previously discussed in this section implicitly scans the region $-1/3 < w_{\rm re} < 1/3$, where $N_{\rm pivot}$ becomes maximum when the contribution of the last term in Eq.~\ref{nppp} is minimum (vanishes) for $w_{\rm re} = 1/3$, i.e. instantaneous reheating. This allows us to put $N^{\rm IRH}_{\rm pivot}$ as the upper bound for $N_{\rm pivot}$ while varying it inside ModeChord+CosmoMC for $-1/3 < w_{\rm re} < 1/3$.
But, in the region $1/3 <w_{\rm re} <1$, the contribution from the last term in Eq.~\ref{nppp} becomes positive which increases the value of $N_{\rm pivot}$ beyond $N^{\rm IRH}_{\rm pivot}$. Therefore, we cannot use the previous prior range for $N_{\rm pivot}$ to analyse for $1/3 <w_{\rm re} <1$.

We note that for the case $1/3 <w_{\rm re} <1$ the last term in Eq.~\ref{nppp} contributes maximum when $w_{\rm re}$ is maximum ($w_{\rm re} = 1$) and $N_{\rm re}$ is maximum also. Now, $N_{\rm re}$ becomes maximum for the lowest allowed reheating temperature that must be above the Big Bang Nucleosynthesis (BBN) bound, namely $T_{\rm re}  > T_{\rm BBN} = 5.1 \spa \rm{MeV}$ \cite{BBN}. Therefore, we have examined the general reheating scenario by simulating with ModeChord+CosmoMC for a few fixed values of $w_{\rm re}$ with the minimum reheating temperature with $T_{\rm re} = T_{\rm BBN}$. For particular values of $w_{\rm re}$ in the range $1/3 <w_{\rm re}< 1$, we set the upper bound on $N_{\rm pivot}$ as $N^{\rm IRH}_{\rm pivot} - {1 \over 4} (1 - 3w_{\rm re})N^{max}_{\rm re} $, where $N^{max}_{\rm re}$ is $N_{\rm re}$ derived with $T_{\rm re} = T_{\rm BBN}= 5.1 {\rm MeV}$ for a fixed value of $w_{\rm re}$. In Sec.~\ref{general_w}, we will discuss in detail how we find $N^{max}_{\rm re}$ for $T_{\rm re} = T_{\rm BBN}$. Our methodology here differs from the previous analyses done in Ref.~\cite{Martin:2014nya, Martin:2013tda,  m3} in terms of parametrisation of the reheating epoch in terms of the underlying model parameters and statistical techniques used for parameter estimation. 

In summary, we incorporate
the effects of reheating using two different methods (and carry out the analysis to obtain the preferred value of the model parameters
using both the methods)
\begin{itemize}
\item[(i)] Analysis using the GRH scenario, and in this case, we vary $N_{\rm pivot}$ between  $N_{\rm pivot}^{\rm IRH}$ as given by \pref{kahler_irh} and 20.
\item[(ii)] Analysis for specific values of $w_{\rm re}$. In this case, $N_{\rm pivot}$ varies between $N_{\rm pivot}^{\rm IRH}$ and the $\hat{N}_{w_{\rm re}}$,
with $\hat{N}_{w_{\rm re}}$, determined by the requirement of reheating temperature being above the temperature needed for successful big bang nucleosynthesis.
\end{itemize}
 
\section{Analysis and results  in the GRH scenario} \label{section_GRH}

  As described above, in the GRH scenario reheating uncertainties are accounted for by varying $N_{\rm pivot}$ between minimum value of 20 and $N^{\rm IRH}_{\rm pivot}$ given by Eq.~\eqref{kahler_irh}. The model parameters (as defined in Sec.~\ref{Seclvs}) are varied in the following ranges: $W_0: 0.001 \spa \textrm{to} \spa130$, $\log_{10} \cv: 5 \spa \textrm{to}  \spa 8$ and $ A_n: 1.80 \spa \textrm{to} \spa 1.95$. We take $g_s =0.06$ (as required for a local realisation of the Standard Model from D3 branes), $\lambda_n = 1$ and  $a_n = 2 \pi$. We keep these parameters fixed as the observables depend mildly on these parameters, and these choices of the parameters are well motivated from the theoretical stand point. Note that among all these parameters $\log_{10} \cv$ also determines the duration of modulus domination epoch, and therefore also affects $N_{\rm pivot}$. The likelihoods used  are \textit{Planck TT+TE+EE}, \textit{Planck lowP}, estimated using commander, \textit{Planck lensing} and \emph{Planck+BICEP2/Keck array} joint analysis likelihood \cite{keck,commander,bkplanck}. 

\begin{figure}[h]
\centering
\includegraphics[width=0.48\textwidth]{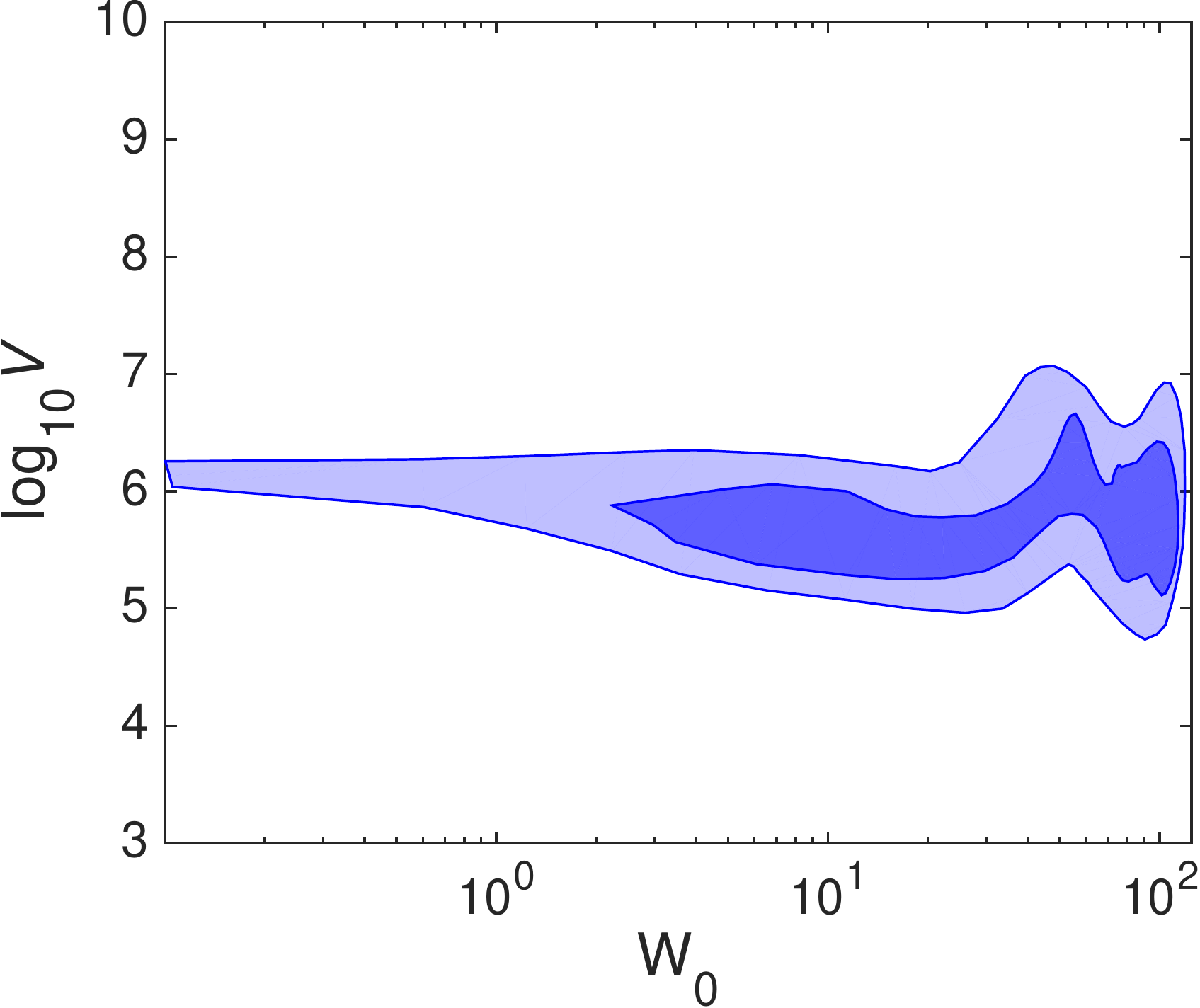}
\caption{Favoured regions in the $W_0$~-~$\log_{10} \cv$ plane. The $1$-$\sigma$ region is shaded as dark blue, the $2$-$\sigma$ region is shaded as light blue.}
\label{fig:f1}
\end{figure}
\begin{figure}[h]
\centering
\includegraphics[width=0.48\textwidth]{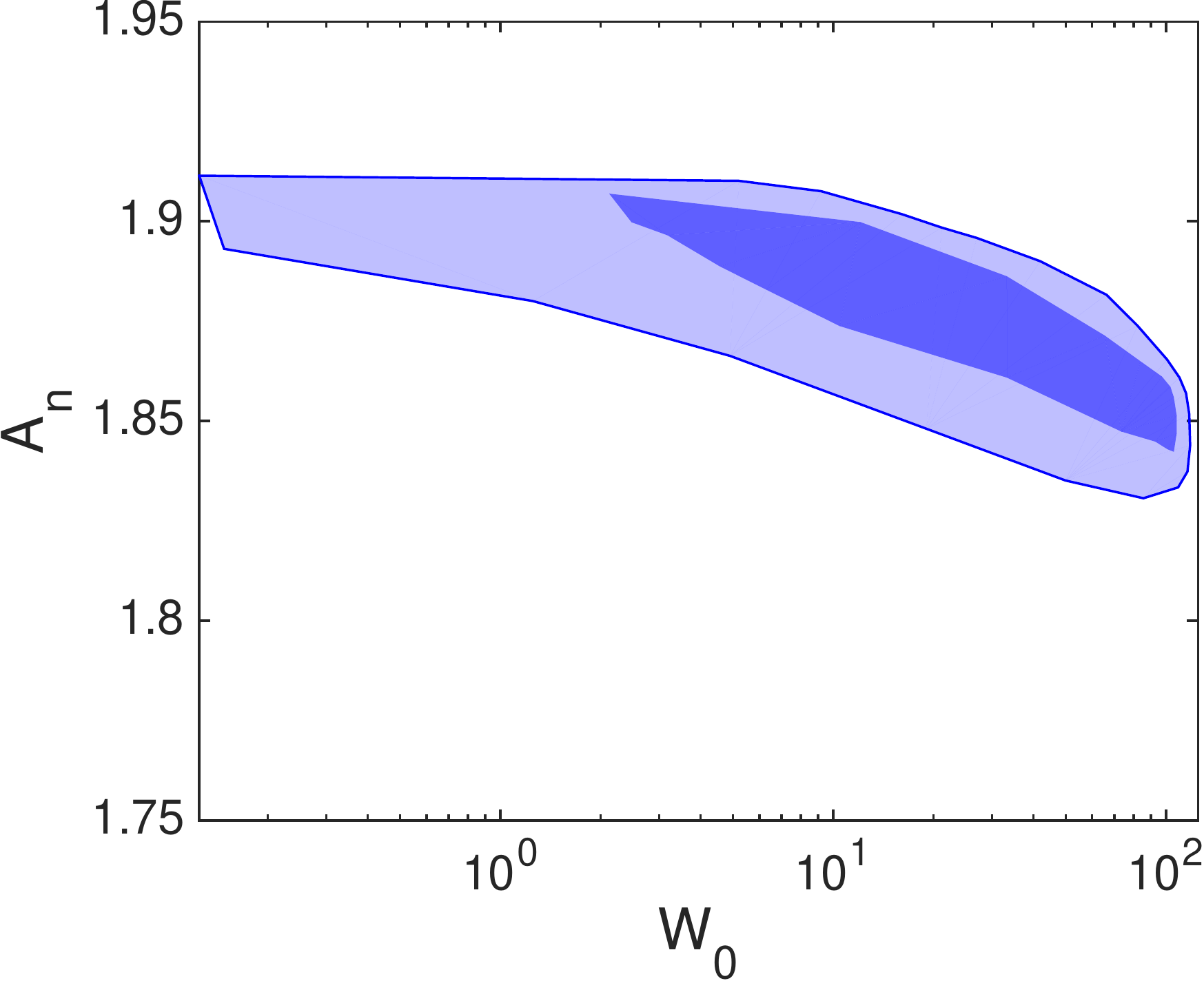}
\caption{Favoured regions in the $W_0$~-~$A_n$ plane. The $1$-$\sigma$ region is shaded as dark blue, the $2$-$\sigma$ region is shaded as light blue.}
\label{fig:f2}
\end{figure}

In Fig.~\ref{fig:f1} and Fig.~\ref{fig:f2}, we show the $1$-$\sigma$ and $2$-$\sigma$ bounds on the model parameters. While Fig.~\ref{fig:f1} shows the marginalised constraint on the parameters $ W_0 $ and $\log_{10} \cv$, Fig.~\ref{fig:f2} shows the marginalised constraint on $W_0$ and $A_n$. These plots represent the most favourable region of the model parameters when $N_{\rm pivot}$ is varied between $20$ to $N^{\rm IRH}_{\rm pivot}$ for this given model. The marginalised central value and the $1$-$\sigma$ errors are quoted in the Table~\ref{table_marge}. Note that $W_0$ is not constrained as tightly as the other two parameters. The central value of the spectral index $n_s \sim 0.953$ obtained from the simulation corresponds to number of $e$-folds $N_{\rm pivot} = 43$. This is in keeping  with the theoretical expectation with $n_s \approx 1- 2/N_{\rm pivot}$, derived under the slow-roll approximations.
\begin{figure}[h]
\includegraphics[width=0.5\textwidth]{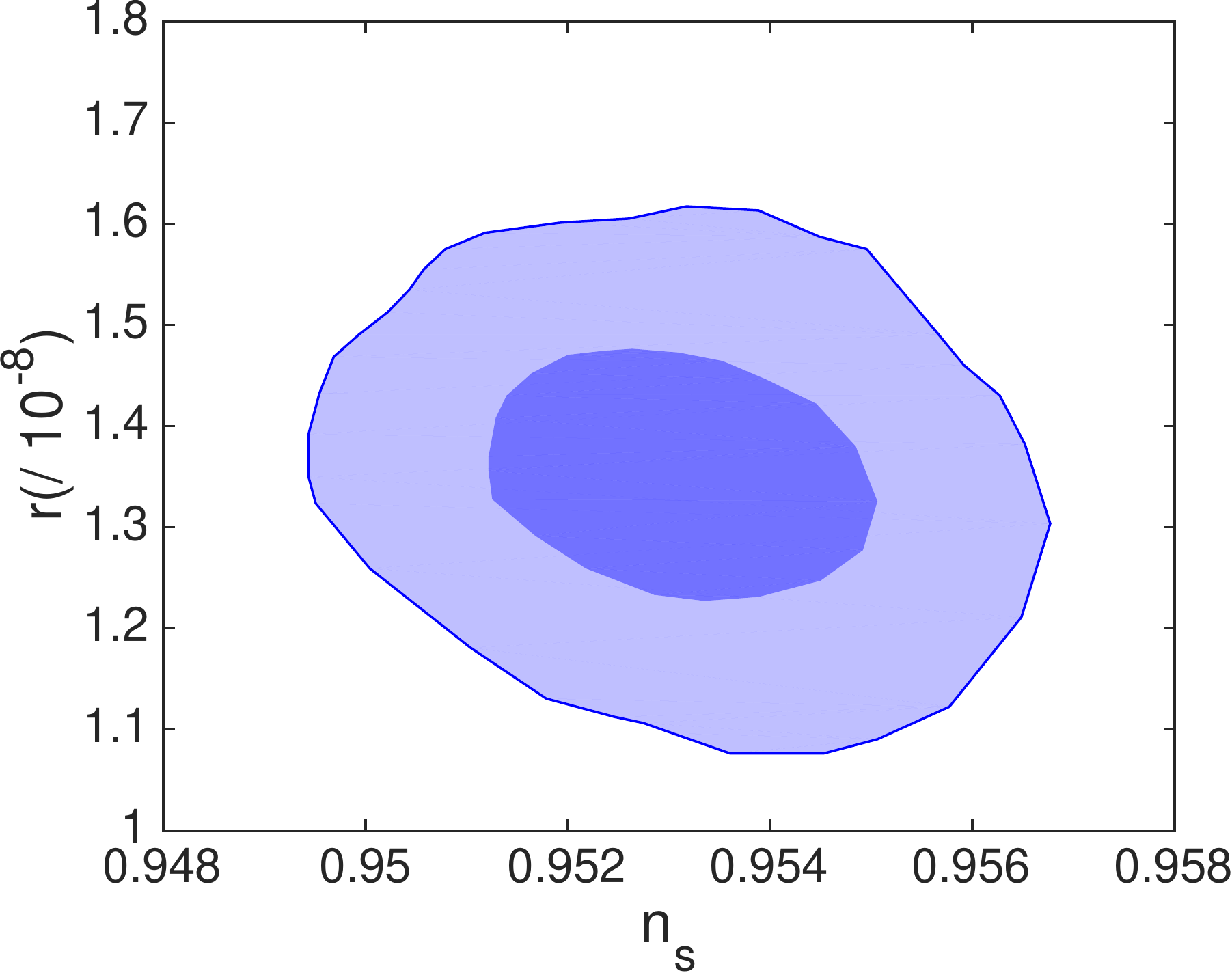}
\caption{Favoured region in the $n_s$-$r$ plane. The $1$-$\sigma$ region is shaded as dark blue, the $2$-$\sigma$ region is shaded as light blue.}
\label{fig:f3}
\end{figure}

\begin{table}[h!]
\centering
\caption{Constraints on the model parameters and the cosmological parameters. Data combination used: $Planck~TT+TE+EE+ low P +lensing + BKPlanck14$.}
\label{table_marge}
\begin{tabular}{|l||l|l|}
\hline
~~~~~Parameters~~~~~&~~~~~Central Value~~~~~&~~~~~$1\sigma$~~~~~\\ \hline \hline
~~~~~~~$W_{0}$~~~~~&~~~~~~~57~~~~~&~~~~~46~~~~~\\ \hline
~~~~~~~$log_{10}\cv$~~~~~&~~~~~~~5.9~~~~~&~~~~~0.3~~~~~\\ \hline
~~~~~~~$A_n$~~~~~&~~~~~~~1.87~~~~~&~~~~~0.04~~~~~\\ \hline
~~~~~~~$n_s$~~~~~&~~~~~~~0.953~~~~~&~~~~~0.002~~~~~\\ \hline
~~~~~~~$r / 10^{-8}$~~~~~&~~~~~~~1.34~~~~~&~~~~~0.1~~~~~\\ \hline
~~~~~~~$N_{\rm pivot}$~~~~~&~~~~~~~43~~~~~&~~~~~2~~~~~\\ \hline
\end{tabular}
\end{table}

\begin{figure}[h!]
\includegraphics[width=3.5in,clip]{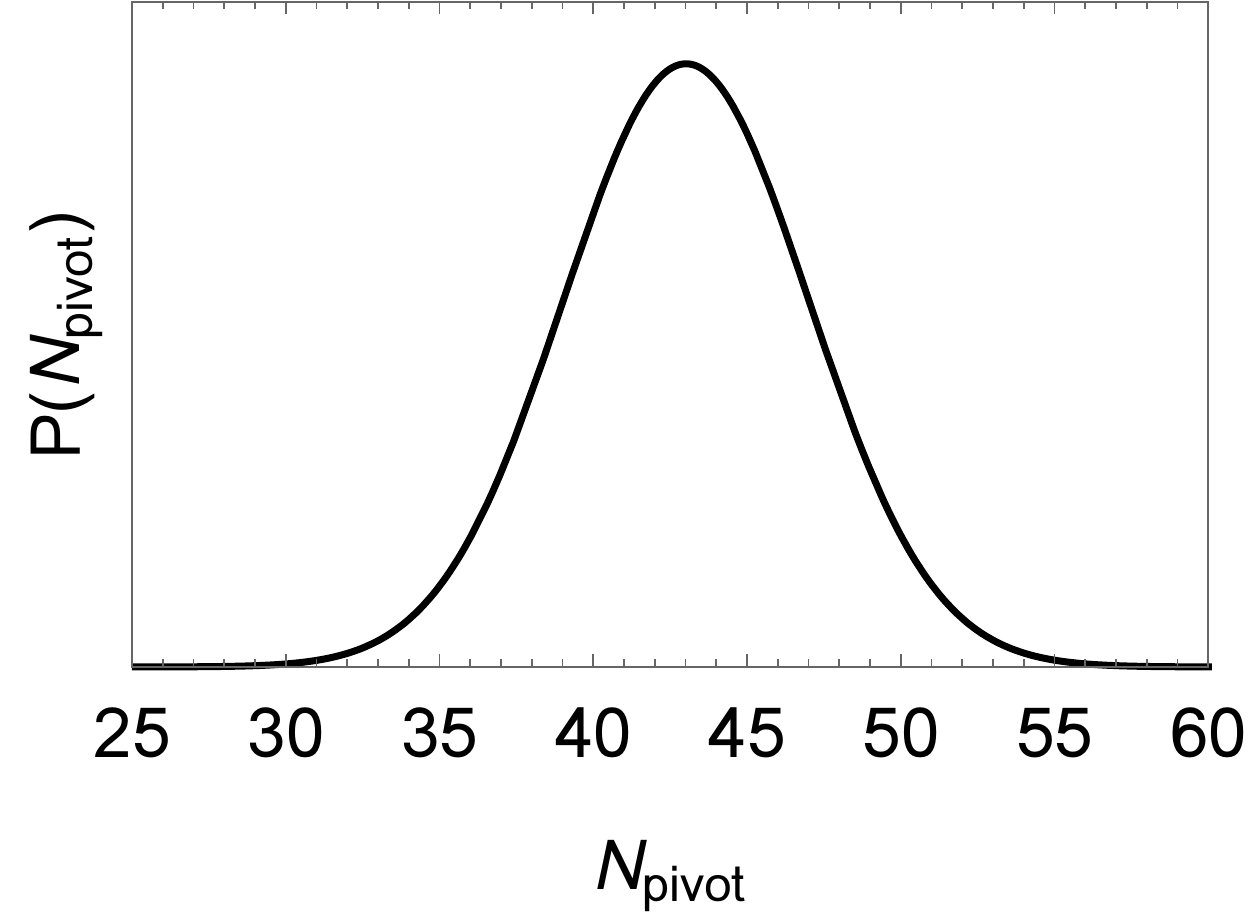}
\caption{1-D probability distribution of the number of e-foldings $N_{pivot}$.}
\label{fig:f4}
\end{figure}

The favoured region in the $n_s$-$r$ plane is presented in Fig.\ref{fig:f3}. Note that the results are in agreement with earlier analytic treatments \cite{BLH, micu}. But, here we would like to emphasise the difference also. In \cite{BLH}, the shift in the $N_{\rm pivot}$ was calculated by using Eq.~\eqref{nmod1} where $\cv \sim 10^5 - 10^6$, fixed by the amplitude of scalar perturbations for typical microscopic parameters. Effectively, the spectral index was calculated at $N_{\rm pivot} \sim 45$ with $n_s \sim 0.955$. But now, we have kept both $\cv$ and $N_{\rm pivot}$ as variables under the generalised reheating scheme, and find preferred values comparing with the data. We present the distribution of $N_{\rm pivot}$ (marginalised over all other parameters) in Fig.~\ref{fig:f4}. We see that as an effect of precision analysis to determine exact values of the model parameters, the central value of $N_{\rm pivot}$ shifts to $43$ from $45$ as found in \cite{BLH}. 

Note that the lower $2$-$\sigma$ bound on $N_{\rm pivot}$ is 39, which is well above 20 and closer to $N^{\rm IRH}_{\rm pivot}=45$. 
 As is evident from the best fit value of $n_s =  0.953\pm 0.002$, the model is outside of the PLANCK ($\Lambda$CDM+$r$) $2$-$\sigma$ lower limit \cite{planck2015}\footnote{Although the model can be consistent when the effects of dark radiation are considered \cite{dark, micu}.}. Our analysis also provides a $\chi^2$ value for the model, and we find that with equal number of parameters to be varied, there is a deterioration of the fit in this case by $\Delta\chi^2 \simeq 13$ with respect to the $\Lambda$CDM+r model for the same combination of the CMB data.
\section{Analysis  and results for specific values of $w_{\rm re}$} \label{general_w}
      In this section we carry out our analysis by making specific choices for $w_{\rm re}$ $(w_{\rm re} = 0, 2/3, 1)$. As discussed earlier,
we will determine the range for variation of $N_{\rm pivot}$ by using the expression for $N^{\rm IRH}_{\rm pivot}$ and the requirement of successful
nucleosynthesis. Before going on to analyse the model for various value of $w_{\rm re}$, let us first describe how we determine this range.

The Hubble parameter at the end of the reheating epoch (after the modulus decay) is given by \cite{BLH}
\begin{align}
H(\hat{t}) = { {M_{\rm pl} W_0^{3}} \over {16 \pi {\cal{V}}^{9/2} (\ln {\cal{V}})^{3/2}} }{ \rm{exp}}\bigg( - { 3 \over 2} (1 + w_{\rm re} ) N_{\rm re} \bigg)~.\label{H(that)}
\end{align}
Moreover, the reheating temperature is given by $ 3 M_{\rm pl}^{2} H^2 ( \hat{t}) = \rho (\hat{t}) \approx {  \pi^{2} \over 30 } g_{*} T_{\rm re}^4$, where $g_{*}$ is the effective number of degrees of freedom of the Standard Model sector. Thus $N_{re}$ can be expressed in terms
of the model parameters, the effective equation of state during reheating and  the reheating temperature:
\bel{nree}
N_{\rm re} = - {2 \over 3}\left( {1 \over {1+w_{\rm re}}}\right)
\ln \left[ \frac{16 \pi^2 g_{*}^{1/2} {\cal{V}}^{9/2} (\ln {\cal{V}})^{3/2} T_{\rm re}^2}{\sqrt{90}M_{\rm pl}^2 W_0^{3}} \right]~.
\ee

\begin{figure}[h!]
\centering
\includegraphics[width=0.48\textwidth]{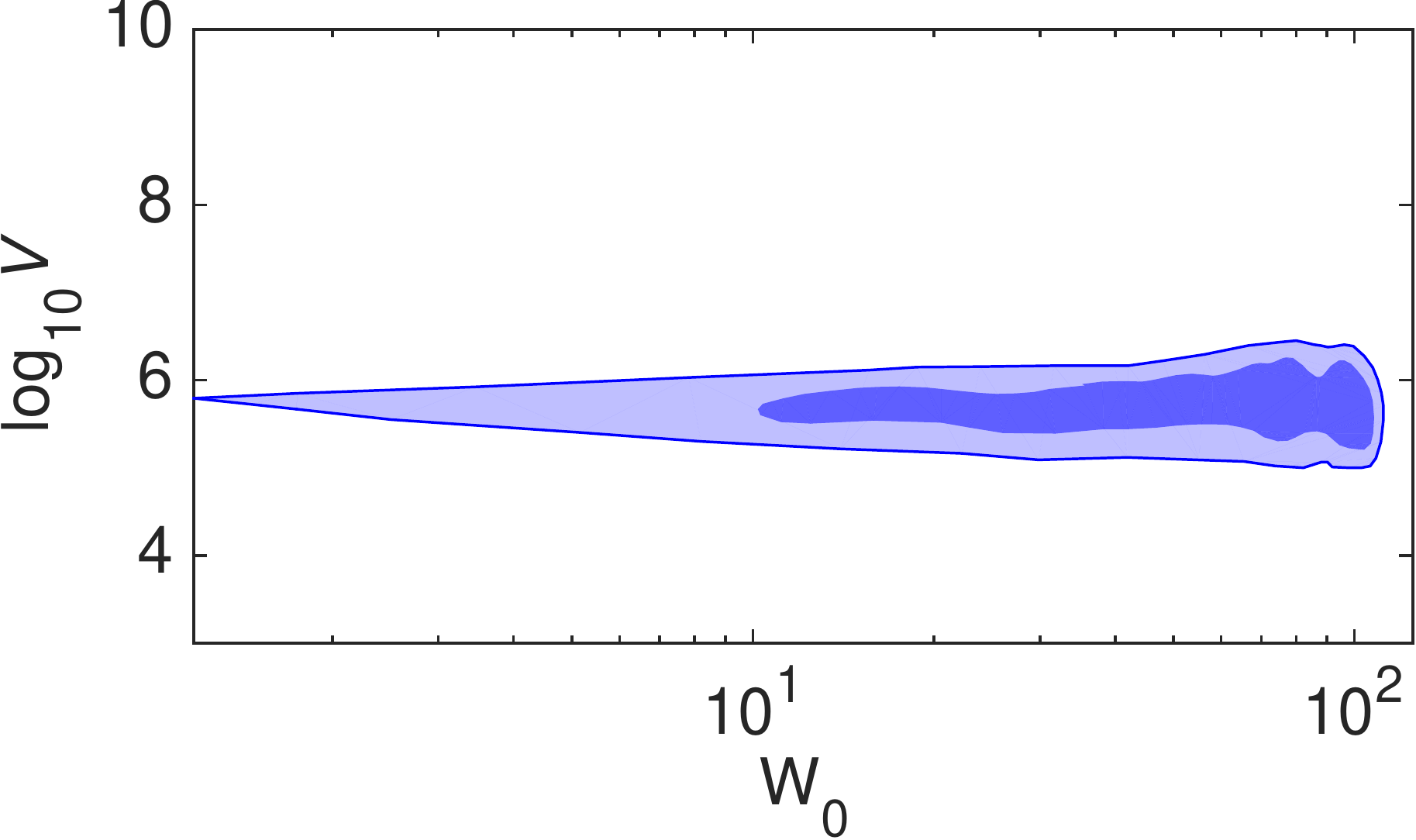}
\caption{Favoured regions in the $W_0$~-~$\log_{10} \cv$ plane for $w_{\rm re} = 2/3$. The $1$-$\sigma$ region is shaded as dark blue, the $2$-$\sigma$ region is shaded as light blue.}
\label{w23_w0logv}
\end{figure}

\begin{figure}[h!]
\centering
\includegraphics[width=0.48\textwidth]{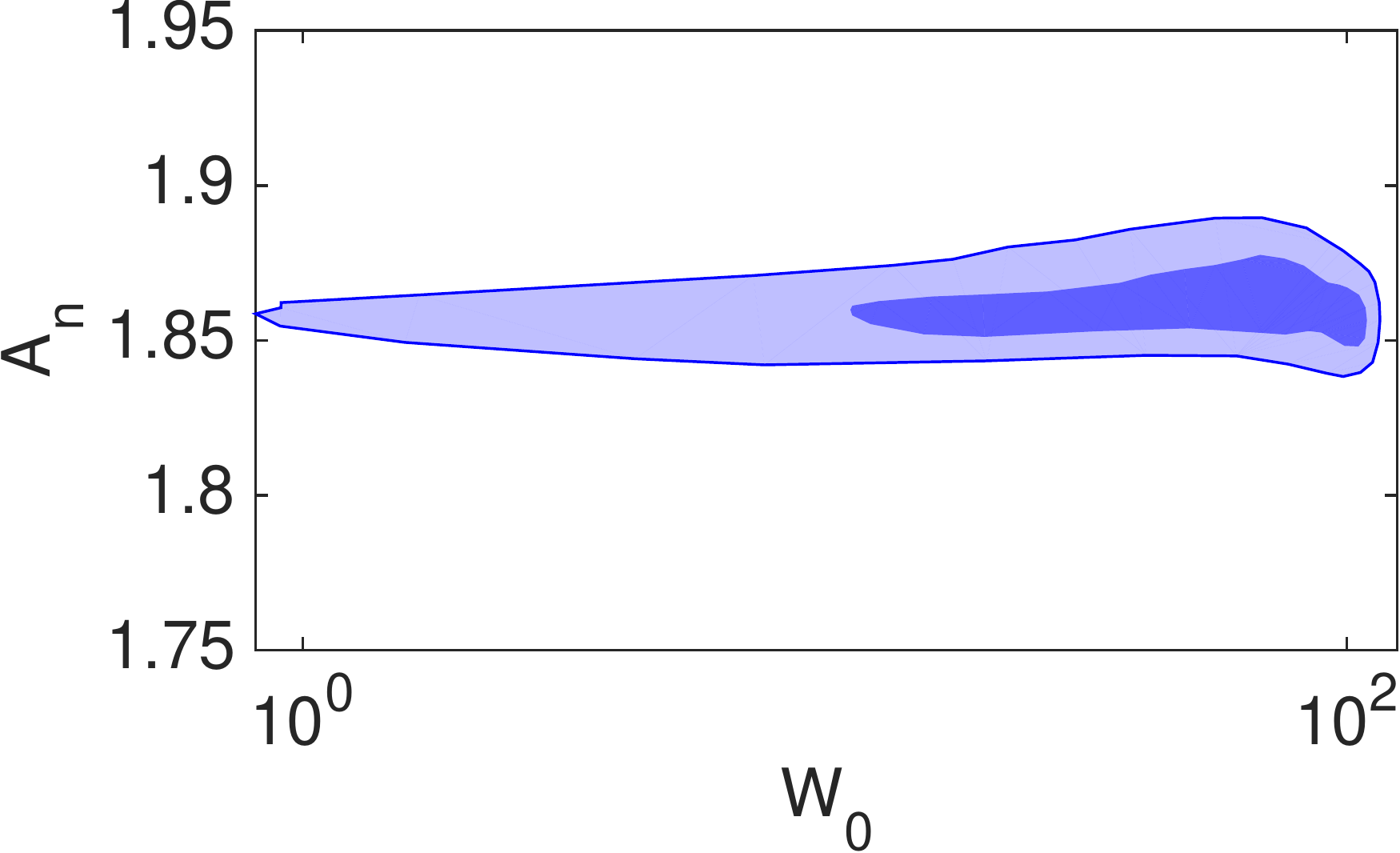}
\caption{Favoured regions in the $W_0$~-~$A_n$ plane for $w_{\rm re} = 2/3$. The $1$-$\sigma$ region is shaded as dark blue, the $2$-$\sigma$ region is shaded as light blue.}
\label{w23_w0An}
\end{figure}

Successful nucleosynthesis requires $T_{\rm re}  > T_{\rm BBN} = 5.1 \spa \rm{MeV}$ \cite{BBN}. Plugging this condition in \pref{nree} we find
an upper bound for $N_{\rm re}$. We will denote this value by $N_{\rm re}^{\rm max}$ (note that this quantity depends on $w_{\rm re}$). Now, in general, 
$N_{\rm pivot}$ is determined by  Eq.~\eqref{nppp}.
%
Since for a given value of $w_{\rm re}$, $N_{\rm re}$ is bounded to lie in the range $(0, N_{\rm re}^{\rm max})$, the allowed range for
$N_{\rm pivot}$ is between $N^{\rm IRH}$ and $N^{\rm IRH} - {1 \over 4} ( 1 - 3 w_{\rm re} ) N_{\rm re}^{\rm max}$. Note that  $N^{\rm IRH} - {1 \over 4} ( 1 - 3 w_{\rm re} )N_{\rm re}^{\rm max}$ is greater than $N^{\rm IRH}$ for $w_{\rm re} > 1/3$, thus for $w_{\rm re} >1/3$, $N_{\rm pivot}$ lies in the interval of 
 $$
 (N^{\rm IRH}, N^{\rm IRH} - {1 \over 4} ( 1 - 3 w_{\rm re} ) N^{\rm max}_{\rm re}).
 $$
\noindent On the other hand for  $w_{\rm re} < 1/3$, $N_{\rm pivot}$ lies in the interval of 
$( N^{\rm IRH} - {1 \over 4} ( 1 - 3 w_{\rm re} ) N^{\rm max}_{\rm re}, N^{\rm IRH})$.
Next, we carry out the analysis to obtain the preferred value of the model parameters for $w_{\rm re} = 1, 2/3,0$.  $N_{\rm pivot}$ is taken to lie within $N^{\rm IRH}$ and $N^{\rm IRH} - {1 \over 4} ( 1 - 3 w_{\rm re} ) N_{\rm re}^{\rm max}$.
\label{exotic_reh}
For all the analyses below, we vary the model parameters $ W_0 $, $\log_{10} \cv$ and $A_n$ in the prior ranges same as section~\ref{results}, i.e., $W_0: 0.001 \spa \textrm{to} \spa130$, $\log_{10} \cv: 5 \spa \textrm{to}  \spa 8$ and $ A_n: 1.80 \spa \textrm{to} \spa 1.95$. The values of the fixed parameters $g_s$ and $a_n$ are also same as section~\ref{results}. For the case of $w_{\rm re} = 2/3$, Fig.~\ref{w23_w0logv} and Fig.~\ref{w23_w0An} are the 2-D marginalised plots for the model parameters, and for $w_{\rm re} = 1$, the plots are similar looking. 

For the sake of completeness, we also do the analysis in this mechanism for a single $w_{\rm re} < 1/3$ case, $w_{\rm re}=0$. Here, the lower bound to $N_{\rm pivot}$ can be specified as $N^{IRH}_{\rm pivot} - {1 \over 4} (1 - 3w_{\rm re})N^{max}_{\rm re}$. Therefore, here, we vary $N_{\rm pivot}$ in the range $N^{IRH}_{\rm pivot} - {1 \over 4} (1 - 3w_{\rm re})N^{max}_{\rm re}< N_{\rm pivot} < N^{IRH}_{\rm pivot} $.


\begin{figure}[h!]
\centering
\includegraphics[width=3.5in,clip]{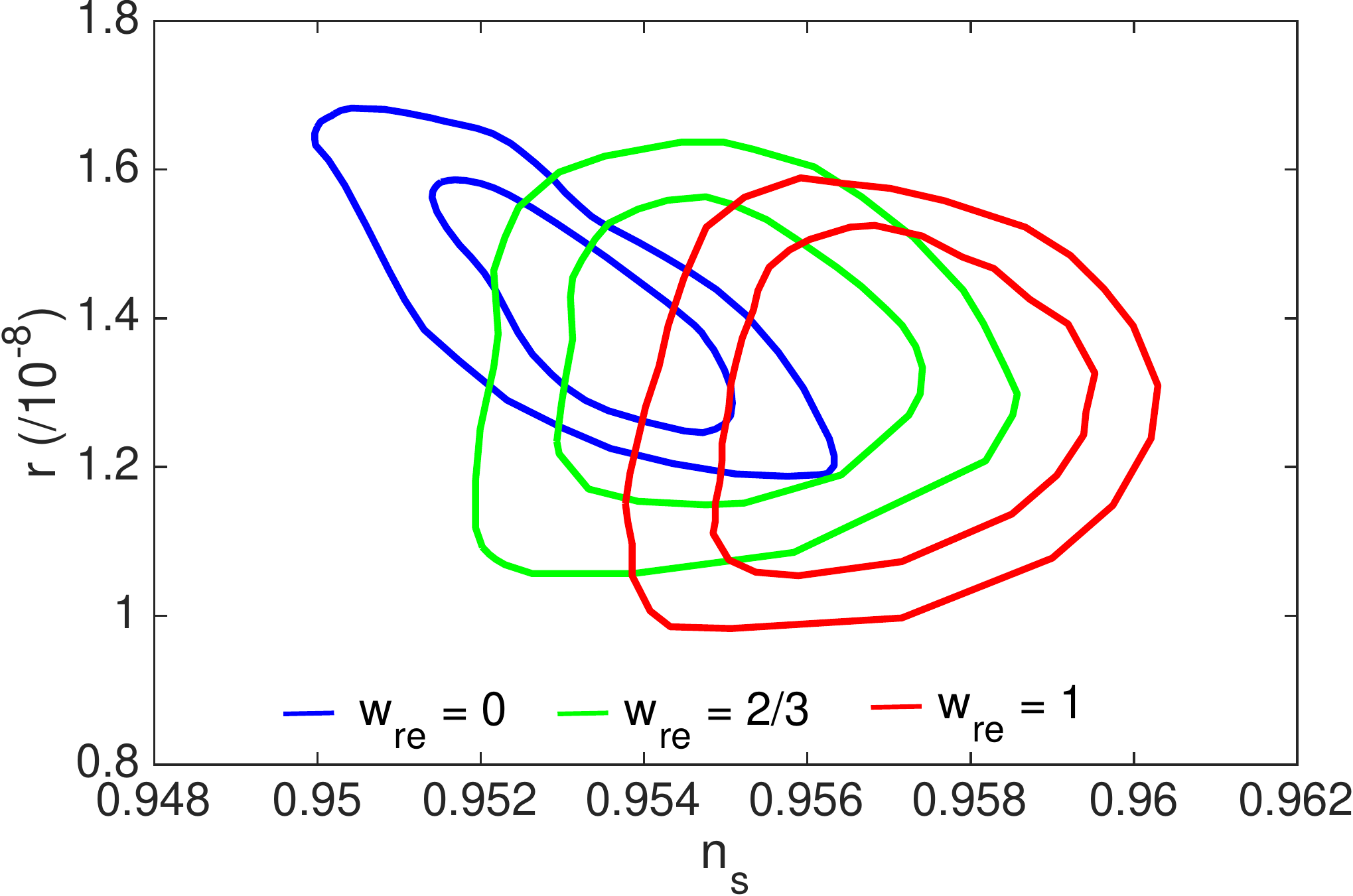}
\caption{$1$-$\sigma$ and $2$-$\sigma$ confidence levels in the $n_s$-$r$ plane for $w_{\rm re} = 0$ (blue contours), $w_{\rm re} = 2/3$ (green contours) and $w_{\rm re} = 1$ (red contours).}
\label{multnsr}
\end{figure}

\begin{figure}[h!]
\includegraphics[width=3.5in,clip]{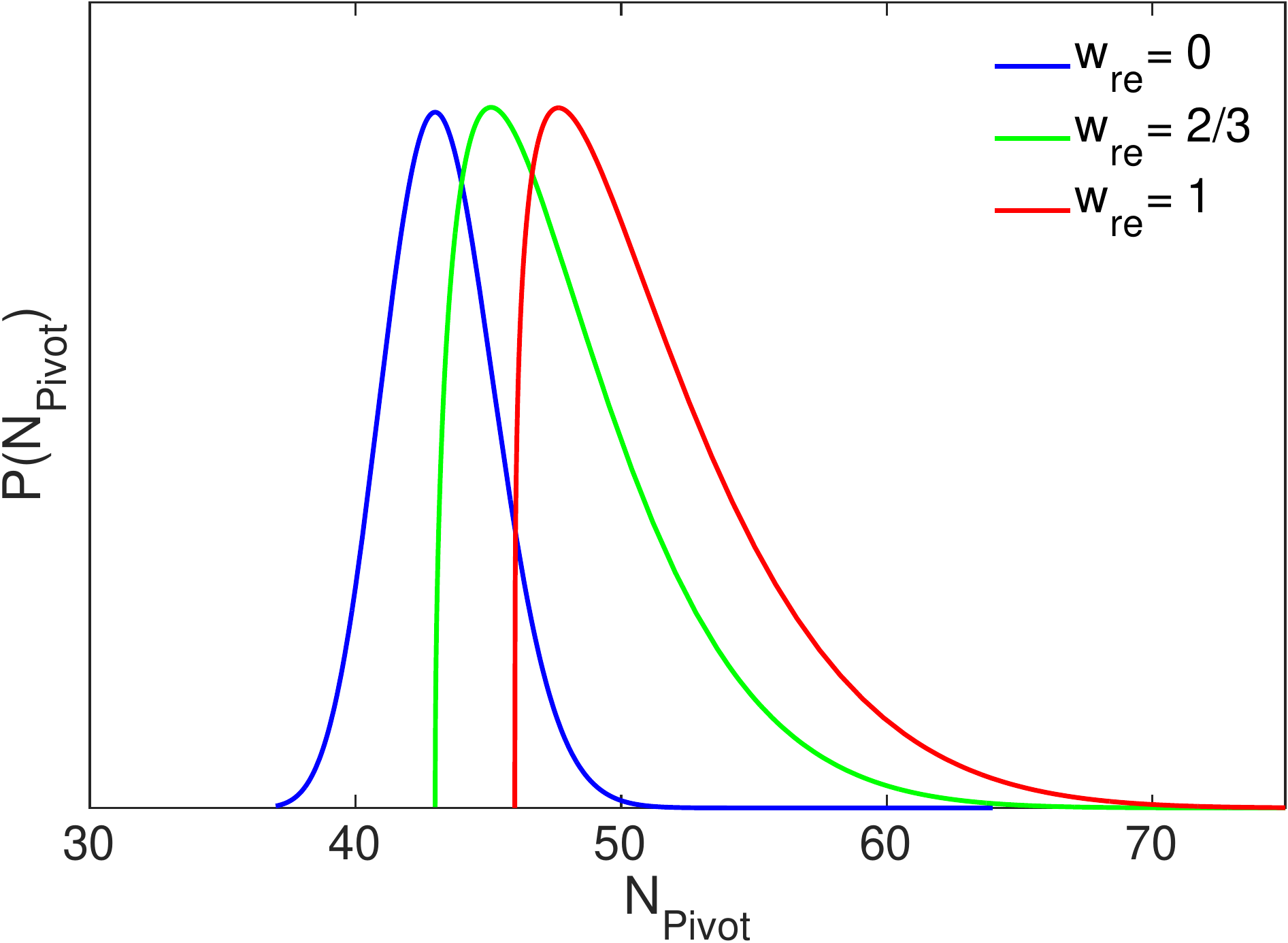}
\caption{1-D posterior probability distribution of the number of e-foldings $N_{\rm pivot}$ for $w_{\rm re} = 0$ (in blue), $w_{\rm re} = 2/3$ (in green) and $w_{\rm re} = 1$ (in red).}
\label{multNpost}
\end{figure}

The best-fit values and 1-$\sigma$ errors for the above three cases $w_{\rm re} = 2/3, 1, 0$ are quoted in Table~\ref{table_marge_multiw}. The values of the model parameters are well within 1-$\sigma$ of the values quoted in Table~\ref{table_marge} in Section~\ref{results}. The 2-D marginalised plot in the $n_s$-$r$ plane is given in Fig.~\ref{multnsr} for the above three cases. The 1-D marginalised posterior distribution for corresponding $N_{\rm pivot}$ are shown in Fig.~\ref{multNpost}.

\begin{table}[h!]
\centering
\caption{Constraints on the model parameters and cosmological parameters for $w_{\rm re} = 2/3, 1, 0$. Data combination used: $Planck~TT+TE+EE+ low P +lensing + BKPlanck14$.}
\label{table_marge_multiw}
\resizebox{0.48\textwidth}{!}{
\begin{tabular}{|l|l|l|l|}
\hline
~~~      ~~~&~~~$w_{\rm re} = 0$~~~&~~~$w_{\rm re} = 2/3$~~~&~~~$w_{\rm re} = 1$~~~\\ \hline \hline
~~~Parameters~~~&~~~Best-fit$\pm 1\sigma$~~~&~~~Best-fit$\pm 1\sigma$~~~&~~~Best-fit$\pm 1\sigma$~~~\\ \hline \hline
~~~$W_{0}$~~~&~~~56.9$\pm$46.5~~~&~~~58$\pm$45~~~&~~~59$\pm$48~~~\\ \hline
~~~$log_{10}\cv$~~~&~~~5.9$\pm$0.3~~~&~~~5.9$\pm$0.3~~~&~~~5.9$\pm$0.3~~~\\ \hline
~~~$A_n$~~~&~~~1.87$\pm$0.04~~~&~~~1.867$\pm$0.03~~~&~~~1.865$\pm$0.05~~~\\ \hline
~~~$n_s$~~~&~~~0.9535$\pm$0.002~~~&~~~0.9555$\pm$0.003~~~&~~~0.9575$\pm$0.003~~~\\ \hline
~~~$r / 10^{-8}$~~~&~~~1.34$\pm$0.1~~~&~~~1.33$\pm$0.1~~~&~~~1.31$\pm$0.1~~~\\ \hline
~~~$N_{\rm pivot}$~~~&~~~43$\pm$2.5~~~&~~~45.2$\pm$2.25~~~&~~~47.7$\pm$2~~~\\ \hline
\end{tabular}}
\end{table}


From table~\ref{table_marge_multiw}, the best-fit values of the the scalar spectral index ($n_s$) for the cases  with $w_{\rm re}>1/3$ is greater than $n_s$ for $w_{\rm re} = 0$ and also greater than the value quoted in Table~\ref{table_marge} for general $w_{\rm re}<1/3$ cases. Moreover, exotic reheating scenarios ($w_{re}> 1/3$) produce $n_s$ values closer to the current marginalised mean values given by Planck 2015 ($\Lambda CDM+r$)~\cite{planck2015} than for the $w_{\rm re}<1/3$ cases. For $w_{\rm re}=2/3$, the value of $n_s$ is just at the lower 2-$\sigma$ bound given by Planck, whereas for the $w_{\rm re}=1$ case, $n_s$ is inside the Planck 2-$\sigma$ bound\footnote{This is consistent to the analysis in Ref~\cite{Bhattacharya2017ysa}.}. Projected sensitivity of $n_s$ in future CMB experiments \cite{Aba} are expected to resolve this situation with stronger constraints. {If we look at the Fig.~\ref{multnsr}, we note that all possible reheating scenarios are consistent to each other at 2-$\sigma$ level. But it is important to appreciate that future observations are going to measure $n_s$ with $\sigma(n_s) \sim 0.002$ at $1$-$\sigma$ level, and in that case, attempts to make meaningful statements about the value of the scalar spectral index automatically requires our better understanding regarding the reheating epoch.} We also note that $N_{\rm pivot}$ has a larger value in the exotic reheating cases, which is expected from the positive contribution of the last term in Eq.~\ref{nppp}. The tensor-to-scalar ratio $r$ is of the same order ($\sim 10^{-8}$) in all of the above cases. 

\bigskip
\section{Conclusions}
\label{conclusions}

In this paper, we have initiated the analysis of string models of inflation using {\sc{ModeChord}}. Given the ultraviolet sensitivity of inflation and the fact that so far the number of inflationary models that have been obtained from string theory is not large \cite{macb}, it is natural to use {\sc ModeChord} when we try to confront them with data. As data becomes more and more precise  $N_{\rm pivot}$ has to be determined very accurately. $N_{\rm pivot}$ itself can explicitly depend on the model parameters for string/supergravity models. Thus, analysis along the line of   the present work will become more pertinent as cosmological observations become more precise. In this work, we constrain model parameters for  K\"ahler moduli inflation, for which the ranges of observables are sensitive to future precision CMB measurements. 

It is known in the literature that an additional post-inflationary era (like moduli domination in our case) is completely degenerate with the re-heating from the CMB point of view \cite{Martin:2010hh} unless the dynamics of the reheating epoch is related to the model parameters\footnote{The degeneracy can also be broken by the detection of primordial gravitational waves, see \cite{Nakayama:2008wy}, \cite{Kuroyanagi:2009br}}.  But this is precisely what happens in the model at hand, the number of e-folds during reheating is known in terms of inflation model parameters which in turn also fixes inflationary observables (this is also the novel feature in the theoretical aspects our analysis of Kahler moduli inflation  in comparison with \cite{Martin:2014nya, Martin:2013tda,  m3}) . We would like to emphasise that the relation between the modulus dominated dominated epoch and the model parameters arose from embedding of the model and our knowledge of the low energy effective action the setting.

There are several interesting directions to pursue. The parameters in the inflationary potentials in string models themselves might have a statistical distribution, and one can try to incorporate the effect of this into the analysis. Another interesting  direction is to understand degeneracies that can arise across the parameters and the model space. It will also be interesting to cross correlate with particle physics observables (see for e.g \cite{Apa, darkexp}) and dark radiation \cite{dark} in LVS.  Note that the constraints of volume and $W_0$ will have direct implications for the supersymmetry breaking scale. The possibility of analysing multi-field models\footnote{For K\"ahler moduli inflation, the single field approximation is valid for a large class of initial conditions \cite{ed}.} can be explored using {\sc{MultiModeCode}.} \cite{mmc}. Another exiting avenue is to develop a better understanding of the reheating epoch\footnote{See \cite{rh,Bhattacharya2017ysa} for a phenomenological approach towards reheating for inflationary models in LVS.} in these models so that associated uncertainties can be reduced.  An recent development in this direction is the possibility that the number of e-foldings during the reheat epoch is bounded \cite{ma}. We hope to return to these questions in near future.

\section*{Acknowledgements}
 Both KD and AM are partially supported by Ramanujan Fellowships funded by SERB, DST, Govt. of India. SB is supported by a fellowship from CSIR, Govt of India. We sincerely thank the referee for  the suggestions on on the initial version of the draft.

\end{document}